\documentclass[showpacs,preprintnumbers,twocolumn,amsmath,amssymb]{revtex4}
\usepackage{graphicx}
\usepackage{dcolumn}
\usepackage{bm}
\usepackage{epsfig}
\usepackage[usenames]{color}

\newcommand{\dhd}{{\textstyle d}
\lower.03ex\hbox{\kern-0.40em$^{\scriptstyle-}$}\kern-0.08em{}}  

\newcommand{\half}{{1\over 2}}

\newcommand{\calj}{{\cal J}}

\begin{document}

\preprint{JLAB-THY-13-1806}

\title{Rapidity evolution of Wilson lines at the next-to-leading order}

\author{Ian Balitsky}
\affiliation{
Physics Dept., Old Dominion University, Norfolk VA 23529,  and\\
Theory Group, Jlab, 12000 Jefferson Ave, Newport News, VA 23606}
\email{balitsky@jlab.org}

\author{
Giovanni A. Chirilli}
\address{
Department of Physics, The Ohio State University,
  Columbus, OH 43210, USA}
\email{chirilli.1@asc.ohio-state.edu}
\date{\today}

\begin{abstract}
At high energies particles move very fast so the proper degrees of freedom
for the fast gluons moving along the straight lines are Wilson-line operators
- infinite gauge factors ordered along the line. In the framework of  operator expansion in Wilson lines 
the energy dependence of the amplitudes 
 is determined by the rapidity evolution of Wilson lines.
We present the next-to-leading order hierarchy of the evolution equations for 
Wilson-line operators.

\end{abstract}

\pacs{12.38.Bx,  12.38.Cy}

\keywords{High-energy asymptotics; Evolution of Wilson lines; Conformal invariance}

\maketitle

\section{\label{sec:in}Introduction }

One of the most successful  approaches to high-energy scattering is the operator expansion in Wilson lines. 
(For a review, see Refs. \cite{mobzor,nlolecture}). This approach is based on factorization in rapidity \cite {prd99} and
the cornerstone of the method is the evolution of Wilson-line operators with respect to their rapidity.
The most well-studied part is the evolution of the
``color dipole'' (the trace of two Wilson lines) which has a great number of phenomenological applications. The evolution
of color dipoles is known both in the leading order (the BK equation \cite{npb96,yura})  and in the next-to-leading 
order (NLO) \cite{nlobk,nlobksym} and
the solutions of the  BK with running $\alpha_s$ \cite{prd75,kw}  are widely used for  $pA$ and heavy-ion experiments at LHC and RHIC.
However, recently it was realized that many interesting processes are described by the evolution of more complicated 
operators such as ``color quadrupoles'' (trace of four Wilson lines) \cite{quadru}. To describe such evolution the NLO BK 
must be generalized to the full hierarchy of Wilson-lines evolution which is the topic of the present paper. 
We were following the method of calculation 
developed in Ref. \cite{nlobk} and the results for many diagrams  (with the notable exception
of  ``triple interaction'' diagrams) can be taken from that paper.  In this
letter-type publication we present only the final results for the kernels and leave the details of calculation for future paper(s).

\section{\label{sec:in}High-energy OPE and rapidity factorization}
Consider an arbitrary Feynman diagram for scattering of two particles with momenta $p_A=p_1+{p_A^2\over s}p_2$ and $p_B=p_2+{p_B^2\over s}p_1$ ($p_1^2=p_2^2=0$).
Following standard high-energy OPE logic we introduce the rapidity divide $\eta$ which separates the ``fast'' gluons 
from the ``slow'' ones. 
As a first step, we integrate 
over gluons with rapidities $Y>\eta$ and leave the integration over $Y<\eta$ to be performed afterwards.
It is convenient to use the background field formalism: we integrate over gluons with $\alpha>\sigma=e^\eta$ and leave gluons with $\alpha<\sigma$ as a background field, to
be integrated over later.  Since the rapidities of the background
gluons are very different from the rapidities of gluons in our Feynman diagrams, the background field can be taken in the form of a shock wave due to the Lorentz contraction.
The integrals over gluons with rapidities $Y>\eta$ give the so-called impact factors -coefficients in front of Wilson-line operators with the  upper rapidity cutoff $\eta$ for emitted gluons. The Wilson lines are defined as
\begin{eqnarray}
&&\hspace{-0mm} 
 U^\eta_x~=~{\rm Pexp}\Big[ig\!\int_{-\infty}^\infty\!\! du ~p_1^\mu A^\sigma_\mu(up_1+x_\perp)\Big],
 \nonumber\\
 &&\hspace{-0mm} 
A^\eta_\mu(x)~=~\int\!d^4 k ~\theta(e^\eta-|\alpha_k|)e^{ik\cdot x} A_\mu(k)
\label{cutoff}
\end{eqnarray}
where $\alpha$ is Sudakov variable ($p=\alpha p_1+\beta p_2+p_\perp$).

The result for the amplitude can be written as 
\begin{eqnarray}
&&\hspace{-0mm} 
A(p_A,p_B)~=~
\label{OPE}\\
 &&\hspace{-0mm} \sum I_i(p_A,p_B,z_1,...z_n; \eta) \langle p_B|U^\eta(z_1)....U^{\dagger\eta}(z_n)|p_B\rangle
\nonumber
\end{eqnarray}
where the color indices of Wilson lines are convoluted in a colorless way (and connected by gauge links at infinity). 
As in usual OPE, the coefficient functions (``impact factors'' $I_i$) 
and matrix elements depend on the ``rapidity divide'' $\eta$ but this dependence is cancelled in the sum (\ref{OPE}). It is convenient to
define the impact factors in an energy-independent way (see e.g. \cite{nloif}) so all the energy dependence is shifted to the evolution of 
Wilson lines in the r.h.s. of Eq. (\ref{OPE}) with respect to $\eta$.

To find the evolution equations of these Wilson line operators with respect to rapidity cutoff $\eta$
we again factorize in rapidity.
We consider the matrix element of the set of Wilson lines between (arbitrary) target states and integrate over the 
gluons with rapidity $\eta_1>\eta>\eta_2=\eta_1-\Delta\eta$ leaving the gluons with $\eta<\eta_2$ as
a background field (to be integrated over later).
In the frame of gluons with $\eta\sim\eta_1$ the fields with $\eta<\eta_2$ shrink to a pancake and we obtain  four diagrams of the type shown in Fig. 1.
The result of the evolution of Wilson lines can be presented as infinite hierarchy of evolution equations for 
n-Wilson-line operators. This hierarchy of equations can be constructed from finite number of ``blocks'' 
with this number equal to the order of perturbation theory.  

It should be mentioned that an alternative approach to high-energy scattering in the dense QCD regime 
is to write the rapidity evolution of the wavefunction of the 
target which is governed by the JIMWLK equation \cite{jimwlk}. The one-loop evolution of the JIMWLK Hamiltonian 
summarizes the hierarchy of equations presented  in the next Section. (After completion of this paper we have learned about the paper \cite{nlojimwalk}
 where NLO JIMWLK Hamiltonian is presented.)

\section{LO hierarchy}
In the leading order the hierarchy can be built from self-interaction (evolution of one Wilson line) and 
``pairwise interaction''. The typical diagrams are shown in Fig. 1 
\begin{figure}[htb]
\includegraphics[width=68mm]{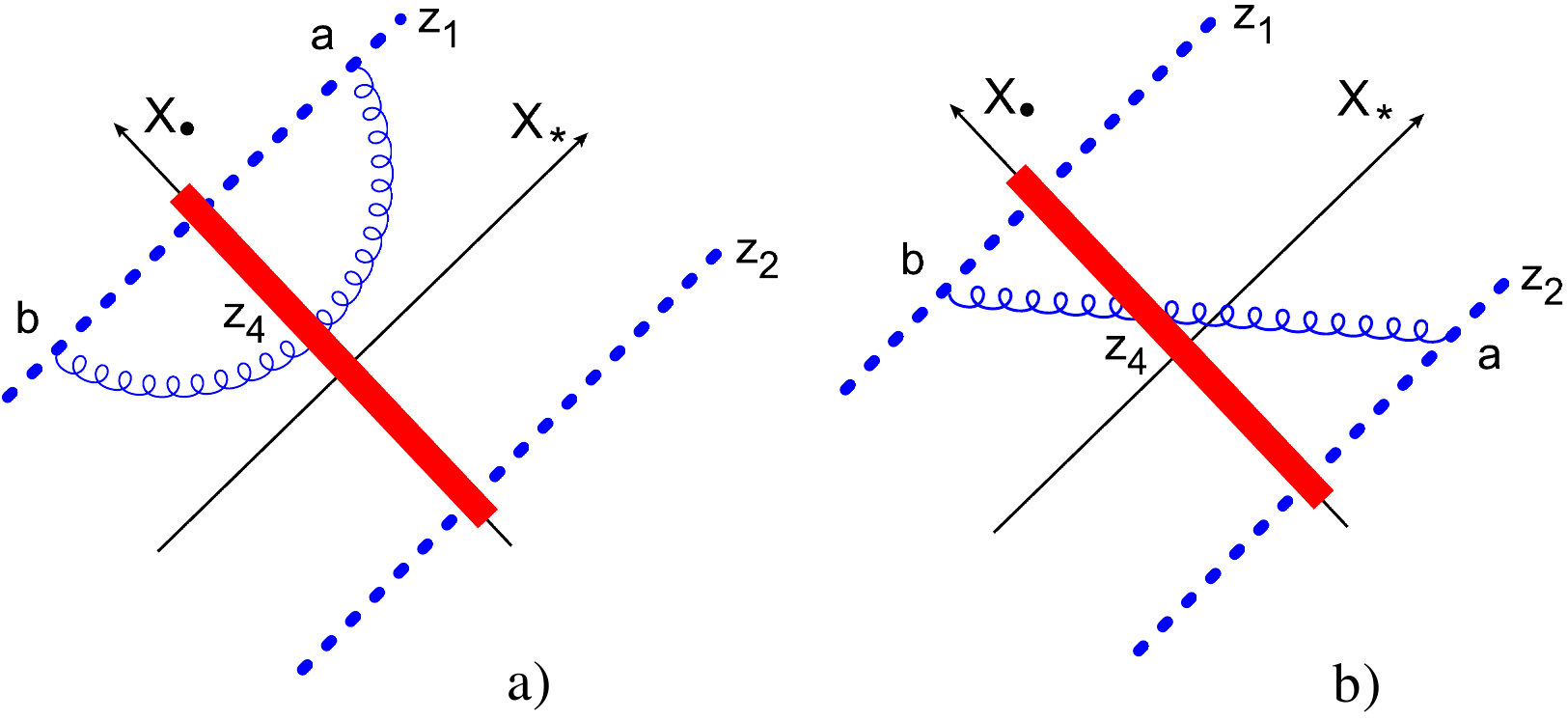}
\caption{LO diagrams.}
\label{fig:lo}
\end{figure}
and the equations have the form \cite{npb96}
\begin{eqnarray}
&&\hspace{-1mm}
 {d\over d\eta}(U_1)_{ij}
~=~
{\alpha_s\over \pi^2}
\!\int {d^2z_4\over z_{14}^2}~(U_4^{ab}-U_1^{ab})(t^aU_1t^b)_{ij}
\nonumber\\
&&\hspace{-1mm}
 {d\over d\eta}(U_1^\dagger)_{ij}
=~
{\alpha_s\over \pi^2}
\!\int {d^2z_4\over z_{14}^2}~(U_4^{ab}-U_1^{ab})(t^bU_1^\dagger t^a)_{ij}
\label{reggeULO}
\end{eqnarray}
for the self-interaction diagrams of Fig. 1a type and 
\begin{eqnarray}
&&\hspace{-1mm}
 {d\over d\eta}(U_1)_{ij}(U_2)_{kl}
~=~
{\alpha_s\over 4\pi^2}\!\int d^2z_4
\big[2U_4 -U_1-U_2\big]^{ab}
\nonumber\\
&&\hspace{-1mm} 
\times~
{(z_{14},z_{24})\over z_{14}^2z_{24}^2}\big[(t^aU_1)_{ij}(U_2t^b)_{kl}+(U_1t^b)_{ij}(t^aU_2)_{kl}\big]
\nonumber\\
&&\hspace{-1mm}
 {d\over d\eta}(U_1)_{ij}(U_2^\dagger)_{kl}
~=~-
{\alpha_s\over 4\pi^2}\!\int d^2z_4
\big[2U_4 -U_1-U_2\big]^{ab}
\nonumber\\
&&\hspace{-1mm} 
\times~
{(z_{14},z_{24})\over z_{14}^2z_{24}^2}\big[(t^aU_1)_{ij}(t^bU_2^\dagger)_{kl}+(U_1t^b)_{ij}(U_2^\dagger t^a)_{kl}\big]
\nonumber\\
&&\hspace{-1mm}
 {d\over d\eta}(U_1^\dagger)_{ij}(U_2^\dagger)_{kl}
~=~
{\alpha_s\over 4\pi^2}\!\int d^2z_4
\big[2U_4 -U_1-U_2\big]^{ab}
\nonumber\\
&&\hspace{-1mm} 
\times~
{(z_{14},z_{24})\over z_{14}^2z_{24}^2}\big[(U_1^\dagger t^a)_{ij}(t^bU_2^\dagger)_{kl}+(t^bU_1^\dagger)_{ij}(U_2^\dagger t^a)_{kl}\big]
\label{2ULO}
\end{eqnarray}
for the ``pairwise'' diagram shown in Fig. 1b. 
Hereafter we use the notation $U_i\equiv U_{z_i}$ and the integration variable 
is called $z_4$ for uniformity of notations in all Sections). All vectors $z_i$ are two-dimensional
and $(z_i,z_j)$ is a scalar product.

The evolution equations in this form are correct both in the fundamental representation of Wilson lines 
where $t^a={\lambda^a/2}$ and in the adjoint representation where $(t^a)_{bc}=-if^{abc}$. In the adjoint
representation $U$ and $U^\dagger$ are effectively the same matrices ($U^\dagger_{ab}=U_{ba}$) so
the three evolution equations (\ref{2ULO}) are obtained from each other by corresponding transpositions.
 (One should remember that $(t^a)_{bc}=-(t^a)_{cb}$ in the adjoint representation). Since the color structure of the
 diagrams in the fundamental representation is fixed one can get the kernels by comparison with adjoint representation.
 Effectively, since our results will be always presented in the form universal for adjoint and fundamental representations
 the NLO results for the evolution of $U\otimes U^\dagger$ and $U^\dagger\otimes U^\dagger$ can be obtained
 by transposition.

\section{NLO hierarchy}

In the next-to-leading order (NLO) the hierarchy can be constructed from
self-interactions,  pairwise interactions, and triple interactions.
The typical diagrams are shown in  Fig. 2 ab, Fig. 2 cd, and Fig. 2 ef, respectively.
\begin{figure}[htb]
\includegraphics[width=55mm]{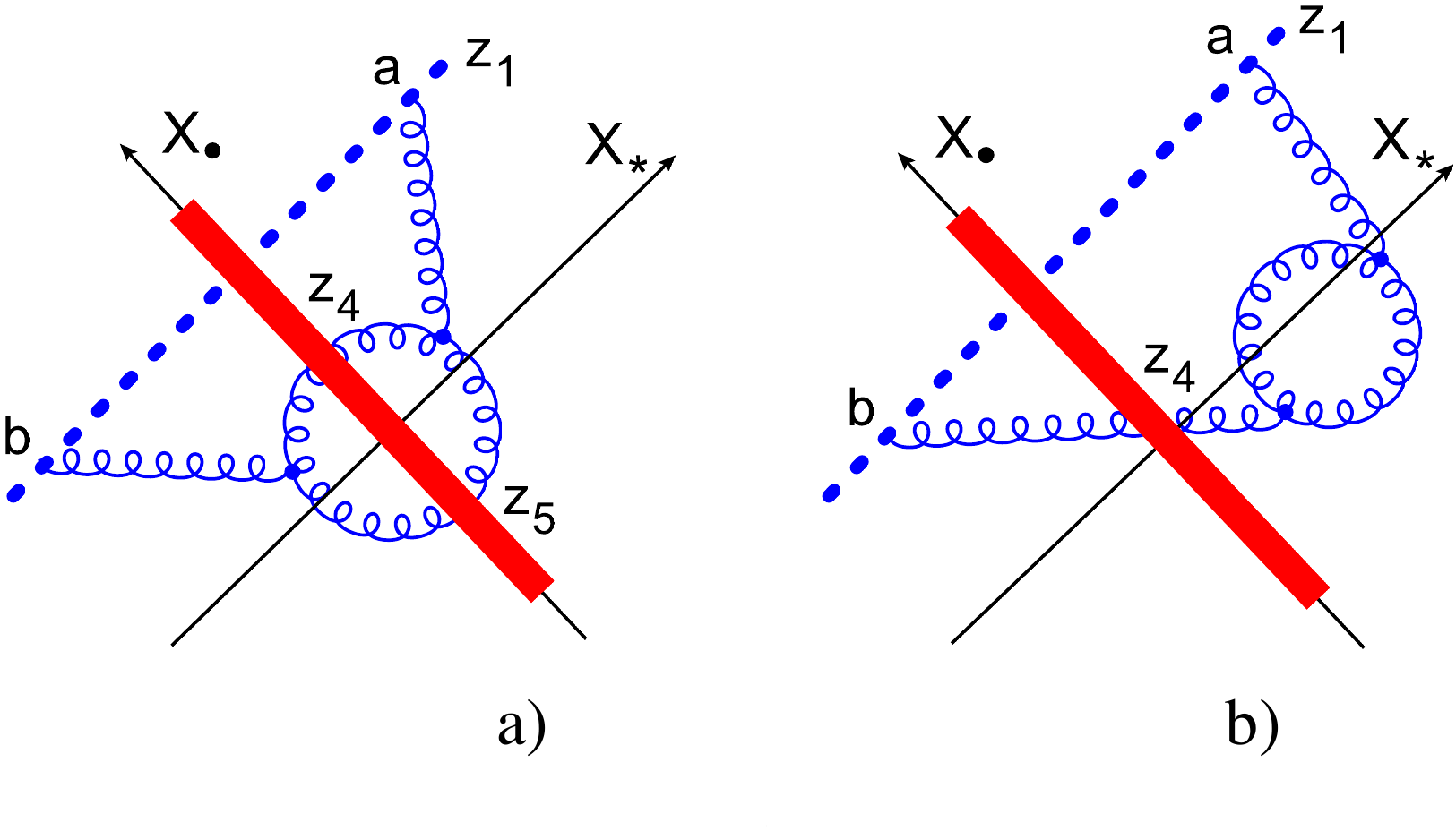}
\vspace{-1mm}
\hspace{-0cm}
\includegraphics[width=66mm]{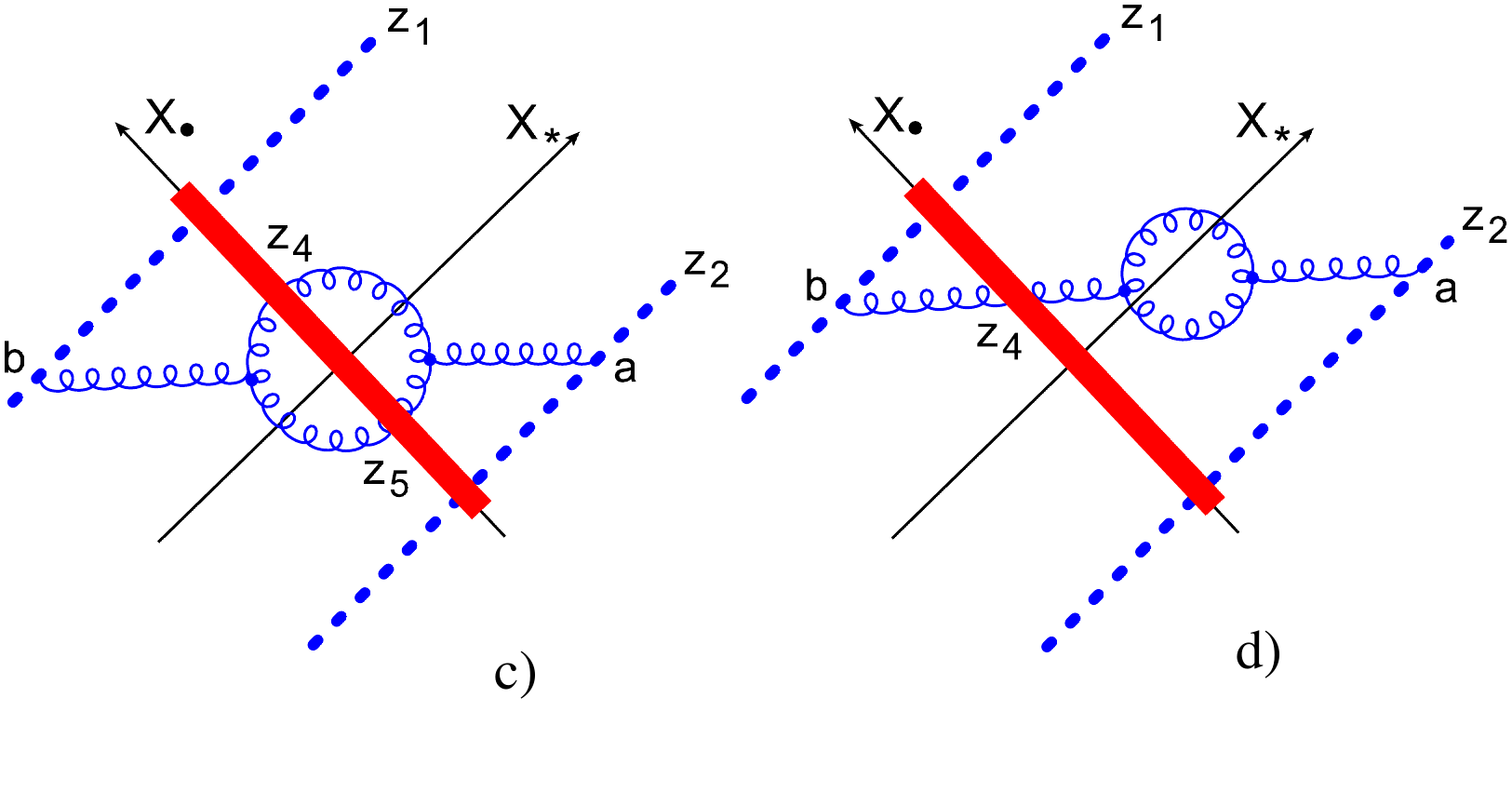}
\vspace{-1mm}
\hspace{-0cm}
\includegraphics[width=66mm]{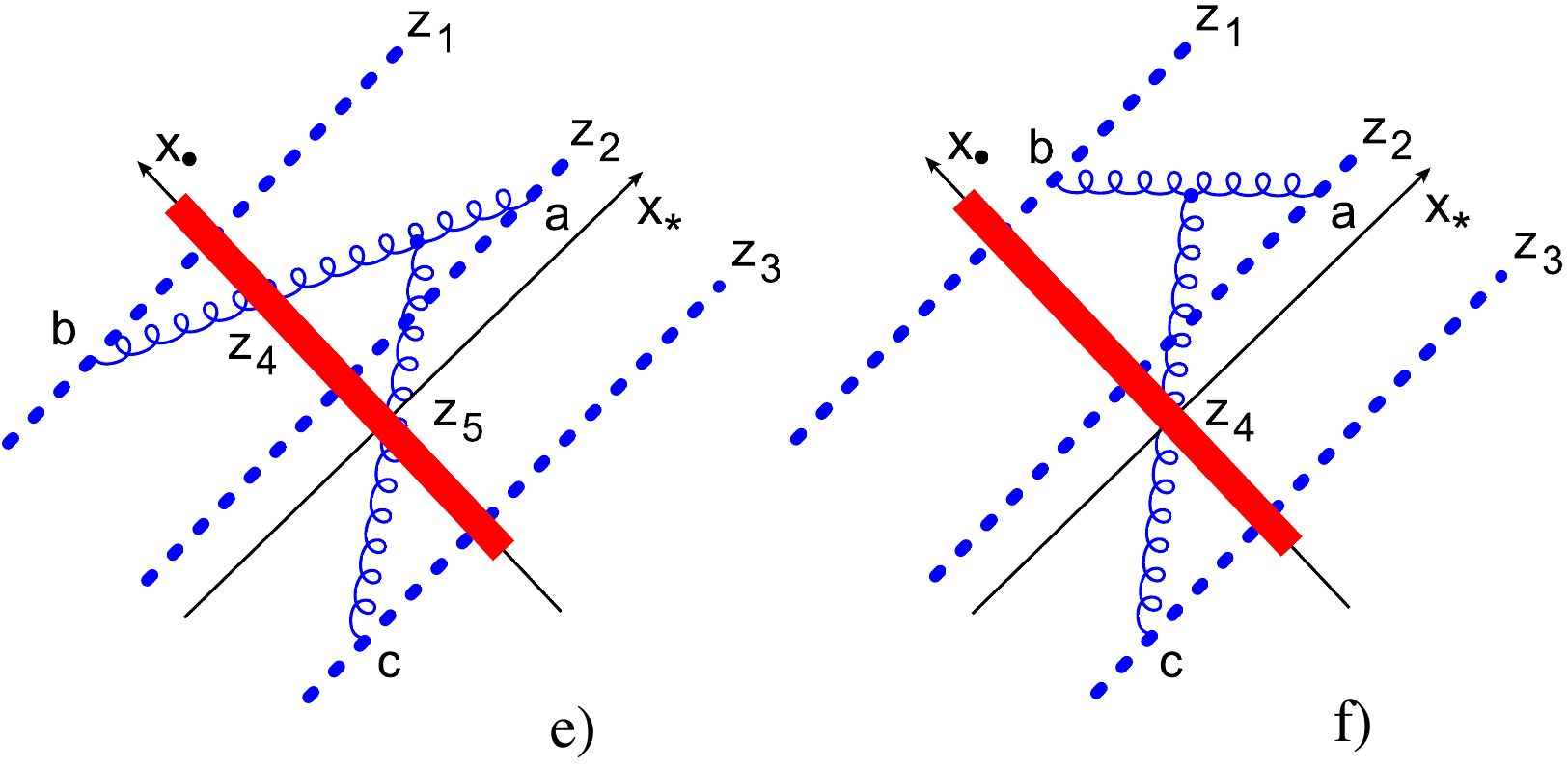}
\caption{Typical NLO diagrams.}
\label{fig:rapidityfac}
\end{figure}

\subsection{Self-interaction}
The most simple part is the one-particle interaction (``gluon reggeization'' term).  The 
typical diagrams are shown in Fig. 2 a,b and the result has the form
\begin{eqnarray}
&&\hspace{-1mm}
 {d\over d\eta}(U_1)_{ij}
=~
{\alpha_s^2\over 8\pi^4}\!\int \!{d^2 z_4d^2 z_5\over z_{45}^2}~\Big\{U_4^{dd'}(U_5^{ee'}-U_4^{ee'})
\nonumber\\
&&\hspace{-1mm}
\times~\Big(\Big[2I_1-{4\over z_{45}^2}\Big]f^{ade}f^{bd'e'}(t^aU_1t^b)_{ij}
+{(z_{14},z_{15})\over z_{14}^2z_{15}^2}\ln{z_{14}^2\over z_{15}^2}
\nonumber\\
&&\hspace{-1mm}
\times~
\big[if^{ad'e'}(\{t^d,t^e\}U_1t^a)_{ij}-if^{ade}(t^aU_1\{t^{d'},t^{e'}\})_{ij}\big]\Big)
\nonumber\\
&&\hspace{-1mm}  
+~8(t^aU_1t^b)_{ij}n_fI_{f1}{\rm tr}\{t^aU_4t^b(U_5^\dagger-U_4^\dagger)\}
\Big\}
\nonumber\\
&&\hspace{-1mm}
+~
{\alpha_s^2N_c\over 4\pi^3}
\!\int\!{d^2z_4\over z_{14}^2}~(U_4^{ab}-U_1^{ab})(t^aU_1t^b)_{ij}
\label{reggeU}\\
&&\hspace{-1mm}
\times~\Big\{\Big[{11\over 3}\ln z_{14}^2\mu^2+{67\over 9}-{\pi^2\over 3}\Big]
-{n_f\over N_c}\Big[{2\over 3}\ln z_{14}^2\mu^2+{10\over 9}\Big]\Big\}
\nonumber
\end{eqnarray}
where $n_f$ is the number of active quark flavors and $\mu$ is the normalization point. 
(The quark diagrams are similar to those in Fig. 2 a-d with the gluon loop
replaced by the quark one). Hereafter we use the notations
\begin{eqnarray}
&&\hspace{-1mm} 
I_1~\equiv~I(z_1,z_4,z_5)
\label{Ix}\\
&&\hspace{-1mm}
=~{\ln z_{14}^2/z_{15}^2\over z_{14}^2-z_{15}^2}\Big[{z_{14}^2+z_{15}^2\over z_{45}^2}
-{(z_{14},z_{15})\over z_{14}^2}-{(z_{14},z_{15})\over z_{15}^2}-2\Big], 
\nonumber
\end{eqnarray}
$I_2\equiv I(z_2,z_4,z_5)$, and
\begin{eqnarray}
&&\hspace{-1mm} 
I_{f1}~\equiv I_f(z_1,z_4,z_5)~=~{2\over z_{45}^2}-{2(z_{14},z_{15})\over z_{14}^2z_{15}^2z_{45}^2}\ln{z_{14}^2\over z_{15}^2}
\end{eqnarray}
 (The integration variables are called $z_4$ and $z_5$ for uniformity of notations
in all Sections). 

The result in this form is correct both in fundamental and adjoint representations.  
(For quark contribution proportional to $n_f$ one should replace $t^a$ by adjoint representation matrices only
in $t^aU_1t^b$ and leave the fundamental $t^a$ and $t^b$ in the quark loop). 
As we discussed in previous Section, this means that the results for the evolution of  $U^\dagger$ can 
be obtained by transposition. We have checked the ``transposing rule'' by explicit calculation.

\subsection{Pairwise interaction}
The typical diagrams for pairwise interaction are shown in Fig. 2 c,d (and the full set  is given by  Fig. 6 in Ref. \cite{nlobk})
In this letter we present the final result, the details would be published elsewhere. The evolution
equation for $U\otimes U$ has the form
\begin{eqnarray}
&&\hspace{-1mm}
{d\over d\eta} (U_1)_{ij}(U_2)_{kl}
~
\nonumber\\
&&\hspace{-1mm}
=~
{\alpha_s^2\over 8\pi^4}\!\int \!d^2 z_4d^2 z_5({\cal A}_1+{\cal A}_2+{\cal A}_3)
\nonumber\\
&&\hspace{-1mm}
+~{\alpha_s^2N_c\over 8\pi^3}\!\int \!d^2 z_4({\cal B}_1+{\cal B}_2)
\label{UU}
\end{eqnarray}
where the kernels ${\cal A}_i(z_1,z_2,z_4,z_5)$ corresponds to diagrams of Fig.2 a,c type
and ${\cal B}_i(z_1,z_2,z_4)$ to Fig.2 b,d type. The explicit expressions are
\begin{eqnarray}
&&\hspace{-1mm}
{\cal A}_1~=~\big[(t^aU_1)_{ij}(U_2t^b)_{kl}+(U_1t^b)_{ij}(t^aU_2)_{kl}\big]
\nonumber\\
&&\hspace{-1mm}
\times~\Big[f^{ade}f^{bd'e'}U_4^{dd'}(U_5^{ee'}-U_4^{ee'})
\nonumber\\
&&\hspace{-1mm} 
\times~\Big(-K-{4\over z_{45}^4}+{I_1\over z_{45}^2}+{I_2\over z_{45}^2}\Big)
\nonumber\\
&&\hspace{-1mm} 
+~4n_f(I_{f1}+I_{f2}+K_f){\rm tr}\{t^aU_4t^b(U_5^\dagger-U_4^\dagger)\}\Big]
\label{A1}
\end{eqnarray}
\begin{eqnarray}
&&\hspace{-1mm}
{\cal A}_2~=~
4(U_4-U_1)^{dd'}(U_5-U_2)^{ee'}
\nonumber\\
&&\hspace{-1mm}
\Big\{i\big[f^{ad'e'}\!(t^dU_1t^a)_{ij}(t^eU_2)_{kl}
\nonumber\\
&&\hspace{-1mm}
-~f^{ade}(t^aU_1t^{d'})_{ij}(U_2t^{e'})_{kl}\big]J_{1245}\ln{z_{14}^2\over z_{15}^2}
\nonumber\\
&&\hspace{-1mm}
+~i\big[f^{ad'e'}\!(t^dU_1)_{ij}(t^eU_2t^a)_{kl}-f^{ade}(U_1t^{d'})_{ij}(t^aU_2t^{e'})_{kl}\big]
\nonumber\\
&&\hspace{-1mm}
\times~J_{2154}\ln{z_{24}^2\over z_{25}^2}\Big\}
\label{A2}
\end{eqnarray}
\begin{eqnarray}
&&\hspace{-2mm}
{\cal A}_3~=~
2U_4^{dd'}
\Big\{i\big[f^{ad'e'}(U_1t^a)_{ij}(t^dt^eU_2)_{kl}
\label{A3}\\
&&\hspace{41mm}
-f^{ade}(t^aU_1)_{ij}(U_2t^{e'}t^{d'})_{kl}\big]
\nonumber\\
&&\hspace{-1mm} 
\times~\Big[\calj_{1245}\ln{z_{14}^2\over z_{15}^2}+(J_{2145}-J_{2154})\ln{z_{24}^2\over z_{25}^2}\Big]
(U_5-U_2)^{ee'}
\nonumber\\
&&\hspace{-2mm} 
+~i\big[f^{ad'e'}\!(t^dt^e U_1)_{ij}(U_2t^a)_{kl}-f^{ade}(U_1t^{e'}t^{d'})_{ij}(t^aU_2)_{kl}\big]
\nonumber\\
&&\hspace{-2mm} 
\times~\Big[\calj_{2145}\ln{z_{24}^2\over z_{25}^2}+(J_{1245}-J_{1254})\ln{z_{14}^2\over z_{15}^2}\Big](U_5-U_1)^{ee'}\Big\}
\nonumber
\end{eqnarray}
for ${\cal A}_i$ kernels and 
\begin{eqnarray}
&&\hspace{-1mm}
{\cal B}_1~=~
2\ln{z_{14}^2\over z_{12}^2}\ln{z_{24}^2\over z_{12}^2} 
\nonumber\\
&&\hspace{-1mm}
\times~
\Big\{(U_4-U_1)^{ab}i\big[f^{bde}(t^aU_1t^d)_{ij}(U_2t^e)_{kl}
\nonumber\\
&&\hspace{-1mm}
+f^{ade}(t^eU_1t^b)_{ij}(t^dU_2)_{kl}\big]
\Big[{(z_{14},z_{24})\over z_{14}^2z_{24}^2}-{1\over z_{14}^2}\Big]
\nonumber\\
&&\hspace{-1mm}
+~(U_4-U_2)^{ab}i\big[f^{bde}(U_1t^e)_{ij}(t^aU_2 t^d)_{kl}
\nonumber\\
&&\hspace{-1mm}
+~f^{ade}(t^dU_1)_{ij}(t^eU_2 t^b)_{kl}\big]\Big[{(z_{14},z_{24})\over z_{14}^2z_{24}^2}-{1\over z_{24}^2}\Big]\Big\}
\label{B1}
\end{eqnarray}
%
\begin{eqnarray}
&&\hspace{-3mm}
{\cal B}_2~=~
\big[2U_4^{ab} -U_1^{ab}-U_2^{ab}\big]
\nonumber\\
&&\hspace{-3mm} 
\Big\{{(z_{14},z_{24})\over z_{14}^2z_{24}^2}
\big[\big({11\over 3}-{2n_f\over 3N_c}\big)\ln z_{12}^2\mu^2+{67\over 9}-{\pi^2\over 3}-{10n_f\over 9N_c}\big]
\nonumber\\
&&\hspace{-3mm} 
+~\big({11\over 3}-{2n_f\over 3N_c}\big)\big({1\over 2z_{14}^2}\ln{z_{24}^2\over z_{12}^2}
+{1\over 2z_{24}^2}\ln{z_{14}^2\over z_{12}^2}\big)
\Big\}
\nonumber\\
&&\hspace{-3mm} 
\times~[(t^aU_1)_{ij}(U_2t^b)_{kl}+(U_1t^b)_{ij}(t^aU_2)_{kl}]
\label{B2}
\end{eqnarray}
for ${\cal B}_i$ kernels.
Here we used the following notations
\begin{eqnarray}
&&\hspace{-1mm}
J_{1245}~\equiv~J(z_1,z_2,z_4,z_5)~=~
\nonumber\\
&&\hspace{-1mm} 
{(z_{14},z_{25})\over z_{14}^2z_{25}^2z_{45}^2}-2{(z_{15},z_{45})(z_{15},z_{25})\over z_{14}^2z_{15}^2z_{25}^2z_{45}^2}
+2{(z_{25},z_{45})\over z_{14}^2z_{25}^2z_{45}^2},~~~~~
\label{J}
\end{eqnarray}
\begin{eqnarray}
&&\hspace{-1mm} 
\calj_{1245}~\equiv~\calj(z_1,z_2,z_4,z_5)~
\nonumber\\
&&\hspace{-1mm}
=~{(z_{24},z_{25})\over z_{24}^2z_{25}^2z_{45}^2}
 -{2(z_{24},z_{45})(z_{15},z_{25})\over z_{24}^2z_{25}^2z_{15}^2z_{45}^2}
\nonumber\\
&&\hspace{-1mm}
+~{2(z_{25},z_{45})(z_{14},z_{24}) \over z_{14}^2z_{24}^2z_{25}^2z_{45}^2}
-2{(z_{14},z_{24})(z_{15},z_{25})\over z_{14}^2z_{15}^2z_{24}^2z_{25}^2}
\label{calJ1}
\end{eqnarray}
\begin{eqnarray}
&&\hspace{-1mm}
K~=~{1\over z_{45}^4}
\Big[{z_{14}^2{z_{25}}^2+{z_{15}}^2z_{24}^2-4z_{12}^2z_{45}^2\over z_{14}^2 z_{25}^2
-z_{15}^2z_{24}^2}\ln{z_{14}^2 z_{25}^2\over z_{15}^2z_{24}^2}-2\Big]
\nonumber\\ 
&&\hspace{-1mm}
+~\half
\Big({z_{12}^4\over z_{14}^2 z_{25}^2-z_{15}^2z_{24}^2}\Big[
{1\over z_{14}^2 z_{25}^2}+{1\over z_{24}^2 z_{15}^2}\Big]
\nonumber\\ 
&&\hspace{-1mm}
+~{z_{12}^2\over z_{45}^2}\Big[{1\over z_{14}^2 z_{25}^2}-{1\over z_{15}^2z_{24}^2}\Big]\Big)
\ln{z_{14}^2 z_{25}^2\over z_{15}^2z_{24}^2}
\label{K}
\end{eqnarray}
and
\begin{equation}
\hspace{-1mm}
K_f~=~{1\over z_{45}^4}
\Big[-2+{z_{14}^2z_{25}^2+z_{15}^2z_{24}^2-z_{12}^2z_{45}^2\over z_{14}^2 z_{25}^2
-z_{15}^2z_{24}^2}\ln{z_{14}^2 z_{25}^2\over z_{15}^2z_{24}^2}\Big]
\label{Kf}
\end{equation}
The conformally invariant kernels $K$ and $K_f$ are parts of the NLO BK equation for dipole evolution.

Again, the result in this form is correct both in fundamental and adjoint representations so  
 the evolution of $U\otimes U^\dagger$ and $U^\dagger\otimes U^\dagger$ can be obtained by transposition of 
 Eqs. (\ref{A1}-\ref{B2}).
If one transposes Wilson line proportional to $U_2$ in the l.h.s and r.h.s. of Eq. (\ref{UU}),  takes trace of Wilson lines  and adds self-interaction terms for $U$ and $U^\dagger$,  
one reproduces after some algebra the NLO BK equation from Ref. \cite{nlobk}. (In doing so one can use the
integral (\ref{masterintegral}) below with replacements $z_3\rightarrow z_1$, $z_1\rightarrow z_2$ so that 
$\calj_{22145}=\calj_{1245}$ and $z_2\rightarrow z_1$, $z_3\rightarrow z_2$ which gives $\calj_{12145}=J_{1245}$.)
It should be noted that, although we calculated all diagrams anew,   the results for two Wilson lines with open indices 
can be restored from the contributions of the individual diagrams  in Ref. \cite{nlobk} since color structure of these diagrams is obvious
even with open indices.

\subsection{Triple interaction}
The diagrams for triple interaction are shown in Fig. 2 e,f (plus permutations). 
The result is
\begin{eqnarray}
&&\hspace{-4mm}
{d\over d\eta}(U_1)_{ij}
(U_2)_{kl}(U_3)_{mn}
\nonumber\\
&&\hspace{-4mm}
=~i{\alpha_s^2\over 2\pi^4}\!\int\!d^2z_4 d^2z_5~\Big\{ \calj_{12345}\ln{z_{34}^2\over z_{35}^2}
\nonumber\\
&&\hspace{-4mm}
\times~f^{cde}\big[(t^aU_1)_{ij} (t^bU_2)_{kl} (U_3t^c)_{mn}(U_4-U_1)^{ad}(U_5-U_2)^{be}
\nonumber\\
&&\hspace{-4mm}
-~(U_1 t^a)_{ij}  (U_2 t^b)_{kl} (t^cU_3)_{mn}(U_4-U_1)^{da}(U_5-U_2)^{eb}\big]
\nonumber\\
&&\hspace{-4mm}
+~\calj_{32145}\ln{z_{14}^2\over z_{15}^2}
\nonumber\\
&&\hspace{-4mm}
\times~f^{ade}\big[(U_1 t^a)_{ij} (t^bU_2)_{kl}  (t^cU_3)_{mn} (U_4-U_3)^{cd}(U_5-U_2)^{be}
\nonumber\\
&&\hspace{-4mm}
-~(t^aU_1)_{ij}\otimes  (U_2 t^b)_{kl}  (U_3 t^c)_{mn}(U_4^{dc}-U_3^{dc})(U_5^{eb}-U_2^{eb})\big]
\nonumber\\
&&\hspace{-4mm}
+~\calj_{13245}\ln{z_{24}^2\over z_{25}^2}
\nonumber\\
&&\hspace{-4mm}
\times~f^{bde}\big[(t^aU_1)_{ij}(U_2 t^b)_{kl}(t^cU_3)_{mn} (U_4-U_1)^{ad}(U_5-U_3)^{ce}
\nonumber\\
&&\hspace{-4mm}
-~(U_1t^a)_{ij}(t^b U_2)_{kl}  (U_3t^c)_{mn}(U_4-U_1)^{da}(U_5-U_3)^{ec}\big]
\label{triple}
\end{eqnarray}
where 
\begin{eqnarray}
&&\hspace{-5mm}
\calj_{12345}~\equiv~\calj(z_1,z_2,z_3,z_4,z_5)~=
-{2(z_{14},z_{34})(z_{25},z_{35})\over z_{14}^2z_{25}^2z_{34}^2z_{35}^2}
\nonumber\\
&&\hspace{-5mm}
-~{2(z_{14},z_{45})(z_{25},z_{35})\over z_{14}^2z_{25}^2z_{35}^2z_{45}^2}
+{2(z_{25},z_{45})(z_{14},z_{34})\over z_{14}^2z_{25}^2z_{34}^2z_{45}^2}
+{(z_{14},z_{25})\over z_{14}^2z_{25}^2z_{45}^2}  
\nonumber\\
\label{calJ2}
\end{eqnarray}
As usual,
the results for the evolution of $U\otimes U\otimes U^\dagger$ etc. can be obtained by transposition of color structures in Eq. (\ref{triple})

The terms with two and one intersections with the shock wave coincide with Ref. \cite{grab}.  When comparing the results for the diagrams
with one intersection (of Fig. 2e type) to that in Ref. \cite{grab} the following integral is useful:
\begin{eqnarray}
&&\hspace{-1mm}
 \!\int\!{d^2z_5\over\pi} \calj_{12345}\ln{z_{34}^2\over z_{35}^2}
 ~\nonumber\\
&&\hspace{-1mm}
=~ \Big\{{(z_{14},z_{24})\over 2z_{14}^2z_{24}^2}
\ln{z_{23}^2\over z_{24}^2}\ln{z_{23}^2\over z_{34}^2}   -z_2\leftrightarrow z_3 \Big\}
\label{masterintegral}\\
&&\hspace{-1mm}
+~\Big\{\Big[{(z_{14},z_{24})(z_{24},z_{34})\over z_{14}^2z_{24}^2}-{(z_{14},z_{34})\over z_{14}^2}\Big]{1\over i\kappa_{23}}
\nonumber\\
&&\hspace{-1mm}
\times~\Big[ {\rm Li}_2\Big({(z_{24},z_{34})+i\kappa_{23}\over z_{24}^2}\Big)-{\rm Li}_2\Big({(z_{24},z_{34})-i\kappa_{23}\over z_{24}^2}\Big)
\nonumber\\
&&\hspace{-1mm}
+~{1\over 2}\ln{z_{24}^2\over z_{34}^2}\ln{(z_{23},z_{24})+i\kappa_{23}\over (z_{23},z_{24})-i\kappa_{23}}\Big]
+z_2\leftrightarrow z_3\Big\}
\nonumber
\end{eqnarray}
where $\kappa_{23}~\equiv~\sqrt{z_{24}^2z_{34}^2-(z_{24},z_{34})^2}$ and ${\rm Li}_2$  is the dilogarithm (which
cancels in the final result (\ref{triple})).

Note that we calculated the evolution of Wilson lines in the light-like gauge $p_2^\mu A_\mu=0$.
To assemble the evolution of colorless operators one needs to combine these equations
and connect Wilson lines by segments at infinity. These gauge links at infinity do not contribute 
to the kernel both in $p_2^\mu  A_\mu=0$ and  Feynman gauge (note, however,  that their contribution is the only non-vanishing one
in $p_1^\mu A_\mu=0$ gauge). Indeed, in the leading order it is easy to see because gluons coming from gauge
links have a restriction $\alpha<e^\eta$ so the gluon connecting points $x,y$ with $x_+=L\rightarrow \infty$ and $z_+=0$ (inside the shockwave) 
will contain the factor $\exp\big(i{p_\perp^2\over\alpha s}L\big)$ which vanishes for $L\rightarrow\infty$ and $\alpha$ restricted from above. 
Similarly one can prove that gauge links at infinity do not contribute to the NLO kernel and therefore the description of the evolution in terms 
of separate Wilson lines in the $p_2^\mu A_\mu=0$ gauge does make sense. 

\section{Conclusion}
 We have calculated the full hierarchy of evolution equations for Wilson-line operators in the next-to-leading approximation.
Two remarks, however,  are in order.

First, our ``building blocks'' for evolution of Wilson lines are calculated at $d=4$ ($d_\perp=2$) so they contain infrared 
divergencies at large $z_4$ and/or $z_5$, even at
the leading order. For the gauge-invariant operators like color dipole or color quadrupole one
can use our $d_\perp=2$ formulas since all these IR divergencies should cancel. If, however, one is 
interested in the evolution of color combinations of Wilson lines (like for octet NLO BFKL \cite{lipatoctet}) some 
of the above kernels should be recalculated in $d=4+\epsilon$ dimensions.

Second,  the NLO evolution equations presented here are ``raw'' evolution equations for Wilson lines with rigid cutoff (\ref{cutoff}).
For example, in ${\cal N}=4$ they lead to evolution equations for color dipole which is non-conformal. The reason 
(discussed in Ref. \cite{nlobksym}) is that the cutoff (\ref{cutoff}) violates conformal invariance so we need an $O(\alpha_s)$ counterterm
to restore our lost symmetry. For the color dipole such counterterm was found in Ref. \cite{nlobksym} and the obtained evolution
for ``composite conformal dipole'' is M\"obius invariant and agrees with NLO BFKL kernel for two-reggeon Green function 
found in Ref \cite{fadin}. Thus, if one wants to use our NLO hierarchy for colorless objects such as quadrupole in ${\cal N}=4$ SYM one should
correct  our rigid-cutoff quadrupole with counterterms which should make the evolution equation for 
``composite conformal quadrupole'' M\"obius invariant. We hope to return to the quadrupole evolution in  future publications.
Another example is the evolution of the three quark Wilson lines $\epsilon_{mnl}\epsilon_{m'n'l'}U_1^{mm'}U_2^{mm'}U_3^{mm'}$ 
(there are both pomeron and odderon contributions to this operator). After subtracting the Ref. \cite{nlobksym} countertems
the NLO evolution equation for this operator becomes  semi-invariant  just as NLO BK in QCD \cite{bagr}. The study is in progress.

\section*{Acknowledgements}
The authors are grateful to A. Grabovsky, H. Weigert and M. Lublinsky for valuable discussions.
This work was supported by contract
 DE-AC05-06OR23177 under which the Jefferson Science Associates, LLC operate the Thomas Jefferson National Accelerator Facility
and by U.S. Department of Energy under Grant No. DE-SC0004286.

 

\section*{References}

\vspace{-5mm}
\begin{thebibliography}{99}

\bibitem{mobzor}
I. Balitsky, {\it ``High-Energy QCD and Wilson Lines''}, 
[hep-ph/0101042] 

\bibitem{nlolecture}
I. Balitsky, 
{\it ``High-Energy Ampltudes in the Next-to-Leading Order''},
arXiv:1004.0057 [hep-ph]

\bibitem{prd99}
I. Balitsky, 
{\it Phys. Rev.} {\bf D60}, 014020 (1999).

\bibitem{npb96}
I. Balitsky, 
{\it Nucl. Phys.}  {\bf B463}, 99 (1996);
{\it ``Operator expansion for diffractive high-energy scattering''},
[hep-ph/9706411]; 

\bibitem{yura}
Yu.V. Kovchegov,  
{\it Phys. Rev.} {\bf D60}, 034008 (1999);
{\it Phys. Rev.} {\bf D61},074018 (2000).

\bibitem{nlobk}
I. Balitsky and G.A. Chirilli,
{\it  Phys.Rev.} {\bf D77}, 014019(2008)

\bibitem{nlobksym}
I. Balitsky and G.A. Chirilli, 
{\it  Nucl. Phys.} {\bf B822}, 45 (2009).


\bibitem{prd75}
I. Balitsky, 
{\it  Phys.Rev.} {\bf D75}, 014001 (2007).

\bibitem{kw}
Yu. V. Kovchegov and H. Weigert,
{\it Nucl. Phys.}  {\bf A784}, 188 (2007);
{\it Nucl.Phys.} {\bf A789}, 260(2007).

\bibitem{quadru}
A. Kovner and M.  Lublinsky,
{\it JHEP} {\bf  0611}, 083 (2006);
A. Kovner and M.  Lublinsky and a Weigert, {\it  Phys.Rev.} {\bf D74}, 114023 (2006);
F. Dominguez,  C. Marquet, A. Stasto, and Bo-Wen Xiao,
 {\it Phys.Rev.} {\bf D87} 3, 034007 (2013),
E. Iancu and D.N. Triantafyllopoulos, {\it JHEP}{\bf 1311}, 067 (2013).

\bibitem{nloif}
I. Balitsky and G.A. Chirilli,
{\it  Phys.Rev.} {\bf D83}, 031502 (2011),
{\it  Phys.Rev.} {\bf D87}, 014013 (2013).
 
\bibitem{jimwlk}
J. Jalilian Marian, A. Kovner, A.Leonidov and H. Weigert,
{\it Nucl. Phys.} {\bf B504}, 415 (1997),
{\it Phys. Rev.} {\bf D59}, 014014 (1999);
J. Jalilian Marian, A. Kovner and H. Weigert,
{\it Phys. Rev.} {\bf D59}, 014015 (1999); 
A. Kovner and J.G. Milhano, {\it Phys. Rev.} {\bf D61}, 014012 (2000);
 A. Kovner, J.G. Milhano and H. Weigert,
{\it Phys. Rev.} {\bf D62}, 114005 (2000); 
H. Weigert, {\it Nucl. Phys.} {\bf A703}, 823 (2002); 
E.Iancu, A. Leonidov and L. McLerran,
{\it Nucl. Phys.} {\bf A692}, 583 (2001), 
{\it Phys. Lett.} {\bf B510}, 133 (2001); 
E. Ferreiro, E. Iancu, A. Leonidov, L. McLerran,
 {\it Nucl. Phys.} {\bf A703}, 489 (2002).

\bibitem{nlojimwalk}
A. Kovner,  M. Lublinsky and Y. Mulian,
{\it Complete JIMWLK Evolution at NLO},\\
e-print  arXiv:1310.0378

\bibitem{grab}
A.V. Grabovsky,
{\it JHEP} {\bf 1309}, 141(2013)


\bibitem{lipatoctet}
V.S. Fadin and L.N. Lipatov,
{\it Phys.Lett.} {\bf B706}, 470 (2012).

\bibitem{fadin}
V.S. Fadin, R. Fiore and A.V. Grabovsky,
{\it Nucl.Phys.} {\bf B831}, 248 (2010).


\bibitem{bagr}
I. Balitsky and A.V. Grabovsky,
in preparation.


\end{thebibliography}
\end{document}